# Dimensionality Decrease Heuristics for NP Complete Problems

April, 10[th], 2011

eduardo.hwang@gmail.com

## Introduction

The vast majority of scientific community believes that P!=NP, with countless supporting arguments. The number of people who believe otherwise probably amounts to as few as those opposing the 2[nd] Law of Thermodynamics. But isn't nature elegant enough, not to resource to brute-force search?

In this article, a novel concept of dimensionality is presented, which may lead to a more efficient class of heuristic implementations to solve NP complete problems. Thus, broadening the universe of man-machine tractable problems.

Dimensionality, as defined here, will be a closer analog of strain energy in nature.

## Development

Take for example a 3-SAT problem instance:

(X+Y+Z).(A+B+C).(Q+W+E).(R+S+T)….

Then, assign the "canonical solution" attempt of all bits equal to 1 (or all equal to zero).

In the first case, the only clauses not satisfied are triple negated conjunctions:

(!A+!S+!D).(!M+!N+!O) ….

The clauses are now grouped in two batches: the right batch with all currently satisfied clauses; and the left batch, with all currently unsatisified clauses.

(!A+!S+!D).(!M+!N+!O) …. //  (X+Y+Z).(A+B+C).(Q+W+E).(R+S+T)….

*Definition: dimensionality (n) is the number of distinct variables that appear on the clauses <u>currently</u> not satisfied (left batch). Here, the word <u>currently</u> is underlined because this is a definition valid for a given iteration (current variable assignment).*

From this setup, one can perform 2^n-1 assignment transitions until a genuine certificate of satisfiability or unsatisfiability is obtained.

***Exemplifying:***

*There are n possible transitions which flip 1 single bit from the n distinct in the current unsatisfiable batch: Combinations amount to C n,1.*

*There are n.(n-1)/2 possible transitions which flip 2 bits from the n distinct in the current unsatisfiable batch: Combinations amount to C n,2.*

*…..until:*

*There is one possible transition which flips all bits from the n distinct in the current unsatisfiable batch: Combinations amount to C n,n.*

To this point, given a high dimensionality, one still has a brute-force search. So the next step is to try variable assignments (from the $2^n-1$ possible) to minimize dimensionality n.

It is clear that flipping the bits of the variable appearing only on the right side does not benefit the dimensionality decreases. At best, the already satisfied clauses remain on the right batch, not increasing the number of distinct bits on the left side.

However, by flipping bits of the left clauses will cause them to be immediately satisified. The only pitfall there being the risk of bringing more clauses from satisfied to the unsatisfied side.

In turn, good heuristics will displace more variables to the right than to the left, resulting in a net gain in dimensionality decrease.

**Example of heuristic: flipping the bit assignment of the most commonly occurring variables on the left sided clauses.** *For very few flippings, many other infrequent variables (sharing clauses with the most frequent) will be automatically be switched to the right side, yielding larger gains in dimensionality decrease.*

**Hardness**: *if the procedure gets stuck at a given dimensionality, many of the $2^n-1$ transitions possess the same level of dimensionality. This is an analog to energy minimization problems, with many (in this case exponential) microstates with similar energy levels, also corresponding to a possible ground energy macrostate.*

At every iteration, the dimensionality is a measure of the maximal operation count (computational distance) to a legit certificate of satisfiability / unsatisfiability. Even for large n (hopefully only at the beginning), all remaining $2^n-1$ transitions are at all times recursively enumerable.

Besides, this dynamic character of the procedure does not a priori "fix" a given variable value until the last iteration. Thus not ruling out any particular solution until the completion certificate is met. This is essentially very different from the DPLL paradigm *[2]*, with fixed binary tree backtrackings.

After completion, a certificate of satisfiability can then be branched to stem more solutions (if they exist).

## *Conclusion*

Dimensionality, as defined here, objectively quantifies the maximum computational effort still needed to a solution certificate for NP complete problems. Possibly being a good analog for strain energy in energy minimization problems found in nature.

Protein Folding is an example of the problem behavior depicted. Levinthal's paradox *[1]* emerges from the exponentially large number of possible molecule conformations, yet resolved very rapidly at small time scales, yielding a correct and functional protein shape. In this case, nature resourced to "funnel-like energy landscapes" rather than brute-force search.

Existing heuristics already in usage may be performing a similar task, with effective dimensionality decrease. However, the precise definition given here allows for a *strict monotonicity* in dimensionality decrease, which is not granted if an algorithm only seeks to minimize the number of unsatisfied clauses.

This subtle difference in heuristic paradigm can add to large amounts of computational time being more efficiently used.

Finally, many other interesting problems which have analogs solved by nature can be made tractable, e.g.: the Steiner Tree problem *[3]*.

## *About the Author*

Eduardo Hwang is a brazilian mechanical engineer, whose work focuses on energy industry and fluid dynamics. Not formally educated in Computational Complexity, but very fascinated by the subject.